\documentclass{webofc}
\usepackage[varg]{txfonts}   % Web of Conferences font
\usepackage{subcaption}
\usepackage{graphicx}

\newcommand{\rT}{\mathbf{r_\perp}}

\begin{document}
%
%\title{Assessing the nuclear structure in Neon and Oxygen}
\title{Small-$x$ structure of oxygen and neon isotopes as seen by the Large Hadron Collider}

\author{\firstname{Pragya } \lastname{Singh}\inst{1}\fnsep\thanks{\email{prasingh@jyu.fi}} \and
        \firstname{Giuliano} \lastname{Giacalone }\inst{2} \and
        \firstname{Bjoern} \lastname{Schenke}\inst{3} \and
        \firstname{Soeren }\lastname{Schlichting}\inst{4}
}

\institute{
Department of Physics, University of Jyvaskyla,  40014, Finland
\and
Institut f\"ur Theoretische Physik, Universit\"at Heidelberg, Philosophenweg 16, 69120, Germany
\and
Physics Department, Brookhaven National Laboratory, Bldg. 510A, Upton, NY 11973, USA
\and 
Fakult\"at f\"ur Physik, Universit\"at Bielefeld, D-33615 Bielefeld, Germany
}

\abstract{%
Results on collisions of $^{16}$O nuclei performed at the Relativistic Heavy Ion Collider (RHIC) have been presented for the first time at Quark Matter 2023 by the STAR collaboration.  $^{16}$O+$^{16}$O collisions are also expected to take place in the near future at the Large Hadron Collider (LHC) at much higher beam energies. We explore the potential of beam-energy-dependent studies for this system to probe small-$x$ dynamics and QCD evolution. We perform 3+1D IP-Glasma simulations to predict the rapidity dependence of the initial geometry of light-ion collisions, focusing on $^{16}$O+$^{16}$O  and $^{20}$Ne+$^{20}$Ne collisions at $\sqrt{s_{\rm NN}} = 70$ GeV and 7 TeV. The choice of $^{20}$Ne is motivated by its strongly elongated geometry, which may respond differently to the effect of the high-energy evolution compared to the more spherical $^{16}$O.  We find that smearing induced by soft gluon production at high energy causes mild variations in the initial-state eccentricities as a function of the collision energy. These effects could be resolved in future experiments and deserve further investigation.
}
\maketitle

\section{Introduction}

Characterizing the quark-gluon plasma (QGP) created in heavy-ion collisions performed at the Relativistic Heavy Ion Collider (RHIC) and the Large Hadron Collider (LHC) requires a good understanding of the initial states of such systems. These can not be computed from first principles, and in theoretical calculations are typically obtained either from phenomenological models or effective theories of the scattering process, such as the color glass condensate effective theory of low-energy QCD \cite{Gelis:2008rw}.

At Quark Matter 2023, the STAR collaboration presented the first results on the collective flow of relativistic collisions of $^{16}$O nuclei at RHIC. As the same collisions are expected to take place in 2024 at LHC at a much higher beam energy, we may soon be able to constrain from experiments a crucial aspect of the initial states of the collisions which remains to date largely unexplored: its beam-energy and rapidity evolution.

In this study, we present preliminary predictions for the beam-energy and rapidity dependence of the initial conditions of light-ion collisions by means of 3+1D IP-Glasma model \cite{Schenke:2016ksl} simulations. For the beam energy, we choose 7 TeV, corresponding roughly to top LHC energy, and 70 GeV, which is close to the top RHIC beam energy, and also corresponds to the centre-of-mass energy of fixed-target ion-gas collisions recorded by the SMOG2 system of the LHCb experiment at the LHC \cite{LHCb:2021ysy}. We perform collisions of both $^{16}$O and $^{20}$Ne nuclei. They are of direct relevance for the experimental campaigns, and enable us to test whether different intrinsic nuclear shapes \cite{Bally:2022vgo,Ryssens:2023fkv} respond differently to the geometric smearing induced by the small-$x$ QCD evolution.

\section{3D IP-Glasma model and event generation} 
Within the CGC, we adhere to the principle of high-energy factorization, wherein the expected value of an inclusive observable, $\mathcal{O}$, at rapidity $y_{obs}$ is given as:
\begin{align}
    \mathcal{O}(y_{obs}) = \int[DU][DV]\mathcal{W}^p_{\Delta y_p}[U]\mathcal{W}^t_{\Delta y_t}[V]\mathcal{O}[U,V]\,.
\end{align}
Here, $\mathcal{W}^{p/t}_{\Delta y_{p/t}}$ represents the weight functional of the projectile/target, formulated in terms of the Wilson lines $U/V$, with rapidity separation $\Delta y_{p/t}=\pm y_{p/t}-y_{obs}$, obtained through the JIMWLK renormalization equation. Refer to \cite{Schenke:2016ksl} for more details and considerations.

\begin{figure}
    \centering
    \includegraphics[width=0.75\textwidth]{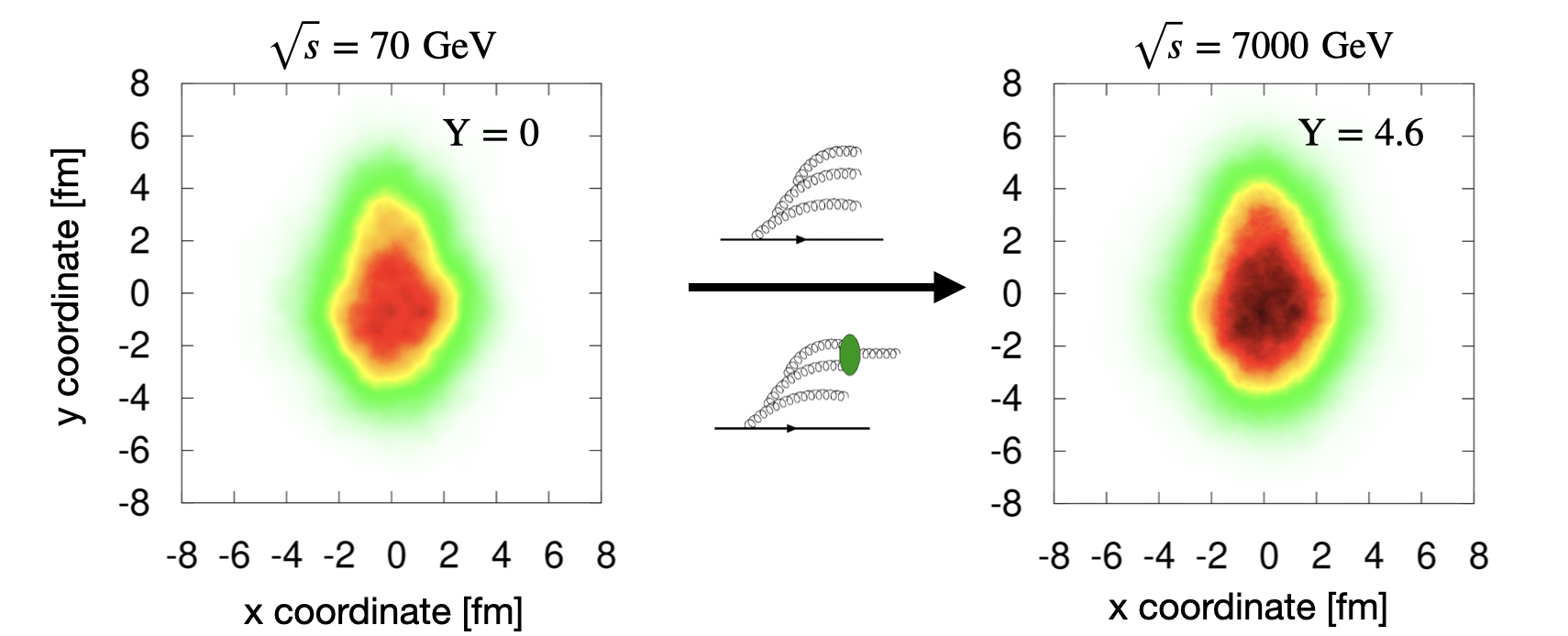}
    \caption{JIMWLK evolution of a particular configuration of $^{20}$Ne nucleus. Shown is $1-\rm{Re~Tr}[V_{\mathbf{x}}]/N_c$ in the transverse plane at rapidities  $Y=0$ (left) and $Y=4.6$ (right).}
    \label{fig:JIMWLK_neon}
\end{figure}

We generate 6k $^{16}$O+$^{16}$O and $^{20}$Ne+$^{20}$Ne collisions. For each colliding nucleus we construct Wilson line configurations computed from nucleon positions sampled on an event-by-event basis. The densities of nucleon centers contain nontrivial spatial correlations, such as clustering and deformations, following the predictions of the \textit{ab initio} Projected Generator Coordinate Method (PGCM) \cite{Frosini:2021sxj}. The 3D IP-Glasma model evolves then the Wilson lines to the relevant beam energies via JIMWLK evolution. In Fig.~\ref{fig:JIMWLK_neon}, we demonstrate the effect of JIMWLK evolution on a particular configuration of $^{20}$Ne. The beam energy can be related to the rapidity, $Y \propto \ln(\sqrt{s}/(70\,{\rm GeV}))$. The small-$x$ evolution weakly smears the nuclear shape, owing to Gribov diffusion as we evolve towards higher rapidity. For each event, we conduct a series of separate 2+1D Classical Yang-Mills simulations at different rapidities to engender a rapidity dependence in the initial geometry. The collisions occur then at various impact parameters. The centrality selection is based on the gluon multiplicity at mid-rapidity ($Y=0$) computed at proper time 0.2 fm/$c$.

% \begin{figure}
%     \centering
%     \includegraphics[width=0.85\textwidth]{PLOTS/chek.jpg}
%     \caption{Geometric eccentricities $\varepsilon_2\{2\}=\sqrt{\langle|\varepsilon_n(y)|^2\rangle}$ as a function of rapidities for different centrality classes for OO (left) and NeNe (right) collision at RHIC (top) and LHC (bottom) energies.}
%     \label{fig:ellipticity}
% \end{figure}

\begin{figure}
  \centering
    \includegraphics[width=0.42\linewidth]{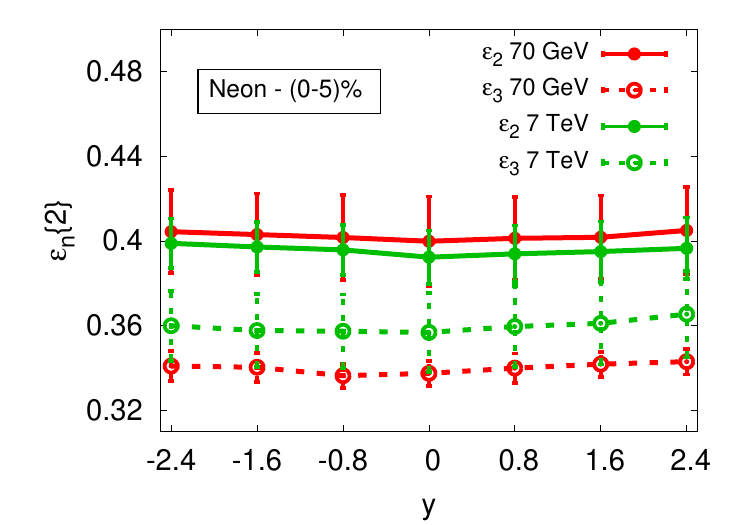}~~~
    \includegraphics[width=0.42\linewidth]{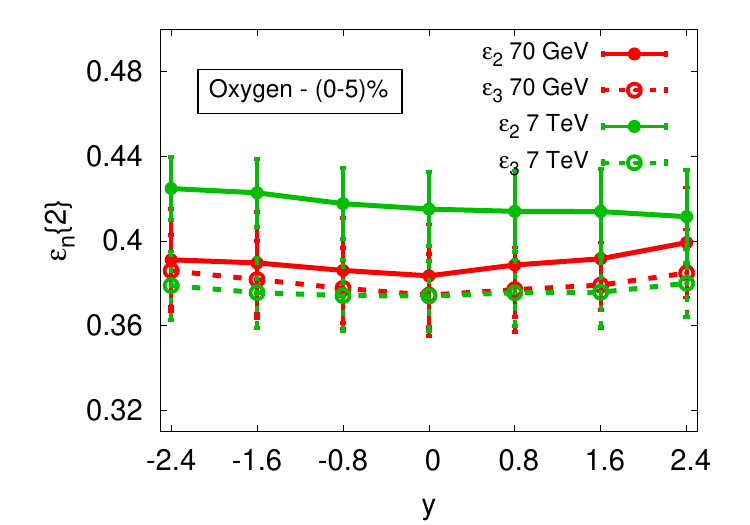}
  \caption{Spatial anisotropies $\varepsilon_n\{2\}=\sqrt{\langle|\varepsilon_n(y)|^2\rangle}$ as a function of rapidity, $y$, for 0-5\% central collisions, at two different energies. Left: $^{20}$Ne+$^{20}$Ne collisions. Right: $^{16}$O+$^{16}$O collisions.} %The other panels depict $\varepsilon_2\{2\}$ as a function of gluon multiplicity for $^{20}$Ne+$^{20}$Ne collisions (mid) and $^{16}$O+$^{16}$O collisions (right). For clarity, the value of $dN_g/dy$ at 70 GeV is rescaled to match that at 7000 GeV in the 0-5\% class.}
  \label{fig:ellipticity}
\end{figure}

\section{Results And Discussion}
We study the energy and rapidity dependence of spatial and momentum eccentricities, i.e.,
\begin{align}\label{eq:en}
        \varepsilon_n(y)=\frac{\int d^2\rT T^{\tau\tau}(y,\rT)~|\rT|^n e^{in\phi_{\rT}}}{\int d^2\rT T^{\tau\tau}(y,\rT)~|\rT|^n}\,, \hspace{5pt}\varepsilon_p(y)=\frac{\int d^2\rT~ T^{xx}(y,\rT)-T^{yy}(y,\rT)+2iT^{xy}(y,\rT)}{\int d^2\rT~T^{xx}(y,\rT)+T^{yy}(y,\rT)}\,.
\end{align}
They are calculated from the glasma stress-energy tensor at $\tau=0.2$ fm/$c$. In the left panel of Fig.~\ref{fig:ellipticity}, we depict $\varepsilon_2(y)$ and $\varepsilon_3(y)$ for the most central $^{20}$Ne+$^{20}$Ne collisions, at both beam energies. The $y$ dependence on these quantities not resolved by our statistical precision. We thus conclude that the rapidity evolution impacts the eccentricities by less than 10\%. More visible seems the impact of the beam energy change, in particular on $\varepsilon_3\{2\}$. This may be of importance for future collisions involving $^{20}$Ne nuclei at LHCb, and deserves further investigation.  Similar conclusions are drawn from  $^{16}$O+$^{16}$O collisions, shown in the right panel of Fig.~\ref{fig:ellipticity}. The results hint at a stronger beam-energy dependence for $\varepsilon_2\{2\}$ in the $^{16}$O case. This may be due to the nuclear geometry. The intrinsic elliptical shape of $^{20}$Ne is hardly affected by the energy evolution (Fig.~\ref{fig:JIMWLK_neon}), possibly leading to an initial-state ellipticity that is preserved across energies. The local fluctuations that dominate instead  $\varepsilon_2$ with more spherical $^{16}$O nuclei may instead be more strongly affected by the small-$x$ evolution, though we do not understand at present the observed increase in $\varepsilon_2\{2\}$ from 70 to 7000 GeV.

%The additional panels in Fig.~\ref{fig:ellipticity} show $\varepsilon_2\{2\}$ as a function of the gluon multiplicity for $^{20}$Ne+$^{20}$Ne collisions (central) and $^{16}$O+$^{16}$O collisions (right) at mid-rapidity. We do not resolve any difference . Conversely, Oxygen, with its tetrahedral configuration, exhibits a substantial increase in ellipticity at LHC energy. Furthermore, a noteworthy observation is the general decrease in ellipticity with increasing multiplicity, except for the most central collision in Oxygen, which warrants further investigation in future studies.

% The rapidity dependence of ellipticity $\varepsilon_2(y)$ for various centrality classes in OO (left) and NeNe (right) at RHIC (top) and LHC (bottom) is compactly summarized in Fig.~\ref{fig:ellipticity}. While oxygen and neon display different qualitative behaviors, the rapidity dependence remains mild across all centralities. Neon exhibits a slightly larger $\varepsilon_2$ in the most central collisions at RHIC due to its elongated shape, while oxygen has a slightly greater value at LHC due to small-$x$ evolution destroying neon's bowling pin shape. Peripheral classes in neon show similar values at both RHIC and LHC, suggesting a fluctuation-driven nature. In Oxygen, mid-peripheral classes have similar values, while the most peripheral class is mostly dominated by fluctuations and has a smaller value.

\begin{figure}[t]
    \centering
    \includegraphics[width=0.42\textwidth]{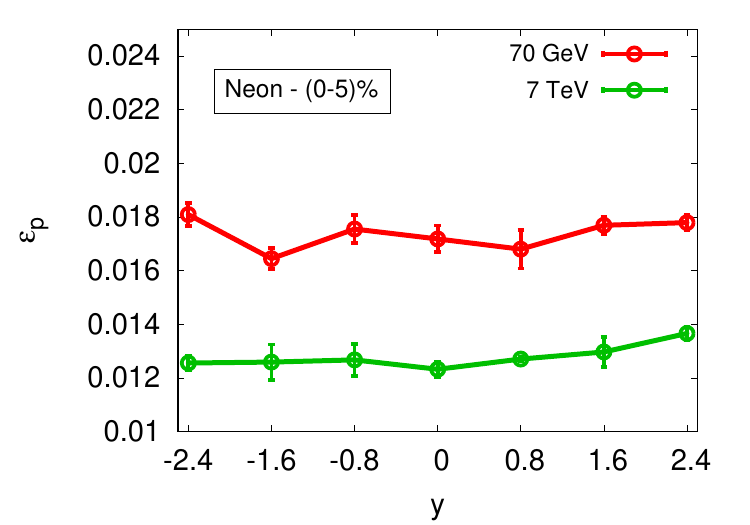}~~ ~
    \includegraphics[width=0.42\textwidth]{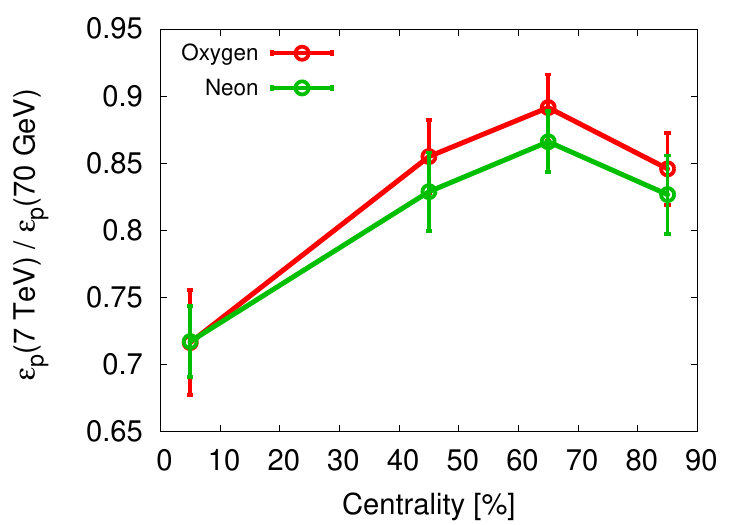}
    \caption{Left: rapidity dependence of initial state momentum anisotropy, $\varepsilon_p(y)$, in 0-5\% central $^{20}$Ne+$^{20}$Ne collisions. Right: centrality-dependent ratio of the mid-rapidity $\varepsilon_p$ taken between the considered energies, for both $^{20}$Ne+$^{20}$Ne and $^{16}$O+$^{16}$O systems.}
    \label{fig:MomentumAnisotropy}
\end{figure}

In Fig.~\ref{fig:MomentumAnisotropy}, the root mean squared value of the initial state momentum anisotropy is shown as a function of rapidity for $^{20}$Ne+$^{20}$Ne collisions at two different energies. As expected, the value of $\varepsilon_p$ is higher at lower beam energies, as it gets reduced by the randomization of color fields induced by the soft gluons emitted at high energy \cite{Schenke:2019pmk}.  The right panel of Fig.~\ref{fig:MomentumAnisotropy} shows the change in $\varepsilon_p$ at different beam energies, evaluated at mid-rapidity, $y=0$, and as a function of the centrality. Peripheral collisions have higher $\varepsilon_p$ due to larger fluctuations in the initial fields. We do not discern any difference between $^{20}$Ne+$^{20}$Ne and $^{16}$O+$^{16}$O collisions.

\section{Conclusion \& Outlook}
We have presented an exploratory study of beam-energy and rapidity-dependent spatial and momentum eccentricities in $^{20}$Ne+$^{20}$Ne and $^{16}$O+$^{16}$O collisions, utilizing the 3+1D IP-Glasma approach. Our observations indicate that the small-$x$ evolution has a mild, though potentially visible impact on the geometry of light-ion collisions. These observations will be substantiated in future by means of higher-statistics calculations. Beyond that, coupling with hydrodynamics and achieving theoretical progress through constructing a fully 3D Wilson line configuration followed by 3+1D CYM \cite{Schlichting:2020wrv,Ipp:2021lwz,Ipp:2017lho,McDonald:2023qwc} would be desirable.

\textit{Acknowledgement}: P.S is supported by the Centre of Excellence in QuarkMatter, project 346324. G.G. and S.S are supported by the Deutsche Forschungsgemeinschaft: Project-ID 273811115
– SFB 1225 ISOQUANT (G.G.), CRC-TR 211 `Strong-interaction matter under extreme conditions’– project
number 315477589 – TRR 211 (S.S). B.P.S. acknowledges support by the U.S. Department of Energy under Contract No. DE-SC0012704 and within the framework of the Saturated Glue (SURGE) Topical Theory Collaboration. The authors wish to acknowledge CSC – IT Center for Science Ltd., Finland, for computational resources on the supercomputer Puhti.

\end{document}